\pdfoutput=1
\documentclass{appolb}

\usepackage{amssymb}
\usepackage{hyperref}
\usepackage{graphicx}
\usepackage{amsmath}
\usepackage{xcolor}
\usepackage{soul}
\usepackage[numbers,sectionbib,sort&compress]{natbib}

\begin{document}

\title{Hadronic form factors in QCD and the incompleteness problem in the time-like region
\thanks{Talk presented by ERA at Excited QCD 2026, 
 8-14 January 2026. Carmen de la Victoria, Universidad de Granada, Spain}
\thanks{Supported by MICIU (Spain) under grant No. PID2023.147072NB.I00 and Junta de Andaluc\'\i a FQM225.}}
\author{Enrique Ruiz Arriola$^1$\thanks{earriola@ugr.es}, 
Pablo Sanchez-Puertas$^{1}$\thanks{pablosanchez@ugr.es}, \\
Wojciech Broniowski$^{2}$\thanks{Wojciech.Broniowski@ifj.edu.pl} 
\address{$^1$Departamento de F\'isica At\'omica, Molecular y Nuclear and \\ Instituto Carlos I de Fisica Te\'orica y Computacional,  Universidad de Granada, E-18071 Granada, Spain}
\address{$^{2}$The H. Niewodnicza\'nski Institute of Nuclear Physics, \\ Polish Academy of Sciences, PL-31342~Cracow, Poland}
}

\maketitle

\begin{abstract}
Hadronic form factors fulfill dispersion relations and
superconvergence sum rules for their spectral density as genuine
imprints of QCD.  We show several instances where these conditions are
flagrantly violated due to the lack of information in the region above
the largest known resonance mass and  below the onset of perturbative QCD. We
propose to use radial Regge trajectories to fill this gap and examine
the consequences of such a ``minimal'' spectral hadronic ansatz. We illustrate the results with the pion charge form factor. 
\end{abstract}



\section{Introduction}

Hadronic form factors (FFs) have been around since Hofstadter proved the finite
proton size in 1955 from the electron scattering experiments; they are identified as
matrix elements of the electromagnetic current. Since then, the nucleon and
other hadronic FFs have been measured or computed in lattice
QCD using all sorts of currents: electromagnetic, axial, gravitational, etc. 
Analyticity, crossing, chiral symmtery, and pQCD play a
key role here and imply a set of normalization and superconvergence sum
rules (SCSRs), which are rigorous conditions. Besides, hadronic FFs
provide an access to the distributions of matter, charge,
currents or stresses inside a hadron. When the currents are conserved,
FFs satisfy normalization conditions. On the other hand,
confinement requires hadronic physical states to be color singlet,
but what is the complete set of eigenstates of QCD spanning the
physical Hilbert space, ${\cal H}_{\rm QCD}$? Moreover, how can we characterize
this completeness property in practice?
In this talk we are concerned with the idea of completeness in hadronic matrix 
elements and their associated FFs, defined such that their
spectral decomposition has definite $J^{PC} I^{G}$ quantum numbers.

\section{Analyticity and sum rules}

Schematically, to simplify the presentation,\footnote{For a
more comprehensive description of the pion and nucleon FFs in
terms of the Lorentz invariant operators see, e.g.,~\cite{Masjuan:2012sk}
and references therein.} we define 
\begin{eqnarray}
\langle H(p')| J | H(p) \rangle = F(q^2) \, , \quad q=p'-p
\end{eqnarray}
where $F(t)$ is the FF and $|H(p)\rangle$ the hadronic
state. For conserved currents, $F(0)$ is usually known from symmetry
requirements and the associated Ward-Takahashi identities.  Due to
crossing, $F(t)$ is analytic in the complex $t$-plane, except for 
branch cuts along the positive real axis above $t>s_{\rm th}$, 
corresponding to the lowest threshold production with suitable quantum numbers.
QCD implies a set of rigurous sum rules based on the asymptotic
behavior in the deep space-like region $q^2 = -Q^2 \to \infty$,
characterized by the running coupling constant and its analytical
extrapolation to the complex plane $q^2 = |s|
e^{i\,{\rm arg} (s)}$, with $ 0 \le {\rm arg} (s) < 2 \pi$:
\begin{eqnarray}
\alpha (s) =  \frac{4\pi/\beta_0}{\log (|s|/\Lambda_{\rm QCD}^2)+ 
i (\arg{s}-\pi)} \Rightarrow \alpha (-Q^2) = \frac{4\pi}{\beta_0} \frac1{\log (Q^2/\Lambda_{\rm QCD}^2)} \, .
\label{eq:as}
\end{eqnarray}
For most cases, one- or two-gluon exchange implies the asymptotics~\cite{Lepage:1979zb,Lepage:1979za}
\begin{eqnarray}
F(s) \sim \begin{cases} \frac{\alpha(s)}{s} \sim \frac1{s(\log s)} \qquad  {\rm mesons}\\
\frac{\alpha(s)^2}{s^2} \sim \frac1{s^2(\log s)^2} \quad {\rm baryons} 
\end{cases} \, . \label{eq:BL}
\end{eqnarray}
Consequently, $ \lim_{|s| \to \infty} F(s) = 0$.\footnote{These are
leading order contributions, whose proportionality constants are, in
general, scale- and scheme-dependent. Higher order contributions are
not well converging for the currently available range in
measurements. Moreover, the onset of pQCD may happen at extremely
large momenta (see e.g. Ref.~\cite{RuizArriola:2008sq,Ananthanarayan:2012tn} for the pion charge FF.)} The point $s=s_{\rm th}$
marks the origin of a branch cut, which goes up to $s \to \infty \pm
i \epsilon$ with a discontinuity ${\rm Disc} F(s) = 2 i {\rm Im}
F(s)$.  Cauchy's theorem applied to a closed contour encircling the
branch cut, yields an {\it unsubtracted} dispersion relation (DR) of
the form
\begin{eqnarray}
F(t)= \frac1{\pi} \int_{s_{\rm th}}^\infty ds \frac{{\rm Im} F(s)}{s-t} \implies F(0) =  \frac1{\pi} \int_{s_{\rm th}}^\infty ds \frac{{\rm Im} F(s)}{s}
\end{eqnarray}
This sum rule implies that the knowledge of the process $J\to
H\bar{H}$ from the lowest (possibly unphysical) threshold up to
infinity allows one to compute $F(0)$ directly.\footnote{This a {\it
prediction} when normalization is not protected by symmetry.
Prominent cases are the magnetic moments or the Druck-term, $D(0)$, of
the nucleon.} In addition, we also have the  SCSRs
\begin{eqnarray}
0 &=&  \frac1{\pi} \int_{s_{\rm th}}^\infty ds~{\rm Im} F(s) \,  \qquad {\rm mesons~and~baryons } \, , \\
0 &=&  \frac1{\pi} \int_{s_{\rm th}}^\infty ds~s~{\rm Im} F(s) \,  \quad {\rm baryons } \, ,
\end{eqnarray}
simply following from the extra suppression with powers of $s$ in
Eq.~(\ref{eq:BL}).  The normalization and SCSR are rigourous theorems
in QCD in the time-like region and direct consequences of the
completeness of hadronic states.  These properties imply that the
spectral functions {\it cannot} be positive-definite in the whole
domain. Still, for the elastic single-channel case, Watson's theorem
implies, $ {\rm Im} F(s) = |F(s)| \sin \delta $ for $ s_{\rm th} \le
s \le s_{\rm in} $, where $s_{\rm in}$ is the threshold for the first
inelastic channel and $\delta$ denotes the phase-shift for the elastic
$\bar H H \to \bar H H$ scattering process.  For an attractive
interaction $\delta >0$ and in this case $\operatorname{Im} F(s) > 0$
below $s_{\rm in}$.  This happens in particular for a resonating
channel, $\bar H H \to R \to \bar H H $.

\section{Violations of the sum rules in practice}

The workings of SCSRs are best appreciated in the case of
the pion charge FF, which has been analyzed in the time-like
region up to $\sqrt{s_{\rm max}}=3$~GeV in terms of a phase-modulus
DR~\cite{RuizArriola:2024gwb} with $F_Q(0)=1$.  From
there, one can extract ${\rm Im}F_Q^\pi(s)$ in the range $4 m_\pi^2 \le
s \le s_{\rm max}$~\cite{Sanchez-Puertas:2024siv,RuizArriola:2025wyq}, 
with ${\rm Im}F_Q^\pi(s) > 0$ for $4 m_\pi^2 \le s \le s_{\rm in}= 4m_K^2$, to find
\begin{eqnarray}
\frac{1}{\pi} \int_{s_{\rm th}}^{s_{\rm max}} ds \frac{\operatorname{Im} F_Q^\pi
  (s)}s  \sim 1 \, , \qquad  
\frac{1}{\pi} \int_{s_{\rm th}}^{s_{\rm max}} ds \operatorname{Im} F_Q^\pi
  (s) \sim m_\rho^2 \, .
\end{eqnarray}
On the other hand, in pQCD to LO one obtains for
$s \to \infty$~\cite{Lepage:1979zb}
\begin{equation}
  \frac1{\pi }{\rm Im} F_Q^{\pi}(s) \to -\frac{16\pi F_{\pi}^2 }{s} {\rm Im }  \alpha_s =
  -\frac{64 \pi^2 F_{\pi}^2 }{\beta_0 \, s} \frac{1}{\log^2 (s/\Lambda_{\rm QCD}^2)+\pi^2} < 0 \, ,
\label{eq:pQCD}
\end{equation}
since ${\rm Im} \alpha_s > 0$
(cf. Ref~\cite{RuizArriola:2025omi} and Eq.~(\ref{eq:as})). Numerically, the contributions to these sum
rules from pQCD are only $-0.0025$ and $-0.114~{\rm GeV}^2$,
respectively, even in the unrealistic scenario where
pQCD extends down to $s = s_{\rm max}$~\cite{Sanchez-Puertas:2024siv,RuizArriola:2025wyq}.
The situation is similar when FFs are deduced from involved 
Roy-Steiner analyses involving the coupled 
$\pi\pi$ and $K \bar K$ channels, and implementing unitarity
and crossing {\it below} the $N \bar N$ threshold, which acts as a high energy
cut-off, $\Lambda= 2 m_N$. For instance, the contributions to SCSRs for the $\pi$ and $K$ gravitational FFs~\cite{Cao:2025dkv}, 
where $s_{\rm th}= s_\pi= 4 m_\pi^2$, is
\begin{eqnarray}
  \frac{1}{\pi} \int_{s_\pi}^{\infty} \mathrm{d} s~\operatorname{Im}A^{\pi,K}(s)=0 \, , \quad \begin{cases}  \frac{1}{\pi} \int_{s_\pi}^{4m_N^2} \mathrm{d} s~\operatorname{Im}A^{\pi}(s)=1.86 {\rm GeV}^2  \\ \frac{1}{\pi} \int_{s_\pi}^{4m_N^2} \mathrm{d} s~\operatorname{Im}A^{K}(s)=0.77 {\rm GeV}^2 
    \end{cases} .
\end{eqnarray}

\begin{table}[ttt]
    \centering
    \caption{Sum rules from the Roy-Steiner analyses for the nucleon~\cite{Hoferichter:2016duk,Cao:2025dkv}. 
    A dash indicates that the value is not provided. }
    \begin{tabular}{l | cc|cc|cc}
    \hline\hline
        & \multicolumn{2}{c}{$n=0$} & \multicolumn{2}{c}{$n=1$} & \multicolumn{2}{c}{$n=2$} \\ \hline
        $\Lambda$ & $2 m_N$ & $\infty$ &   $2 m_N$ & $\infty$  & $2 m_N$ & $\infty$ \\ \hline
        $G_E^v$  & $0.68(11)$ & $1/2$ & - & 0 &- & 0 \\ \hline
       $G_M^v$  & $3.21(30)$ & $2.35$    &-  &0  &-  & 0  \\ \hline
        $A$ & $1.8$ & 1 &   $2.6$ & 0  & $4.0$ & 0  \\ \hline
        $2J$  & $1.4$ & 1& $1.9$ & 0& $2.8$ & 0 \\ \hline
        $\Theta/m_N$ & $2.6$ & 1 & $3.7$ & 0  & $8.1$ & 0 \\
    \hline\hline
    \end{tabular}
    \label{tab.sum rules}
\end{table}

For the nucleon, strong violations occur already for the normalization
($n=0$).  This is illustrated in Table~\ref{tab.sum rules}, which
summarizes a collection of sum rules for the nucleon isovector-vector
electric and magnetic FFs~\cite{Hoferichter:2016duk}, as well as the
gravitational FFs~\cite{Cao:2025dkv}. They correspond to the integrals
\begin{eqnarray}
  S_n (\Lambda) \equiv \frac{1}{\pi} \int_{s_\pi}^{\Lambda^2} \mathrm{d} s~ \frac{\operatorname{Im} F(s)}{s} s^n \, , \qquad n=0,1,2  \, .
\end{eqnarray}
The flagrant violations are due to the finiteness of the upper cut-off
$\Lambda= 2m_N$ and have been mended by introducing effective averaging narrow
resonances, unrelated to the PDG ones, designed to fulfill all sum rules~\cite{Hoferichter:2016duk,Cao:2025dkv}.



\section{Minimal spectral hadronic ansatz}

\begin{figure}
\begin{center}
\includegraphics[angle=0,width=.49\textwidth]{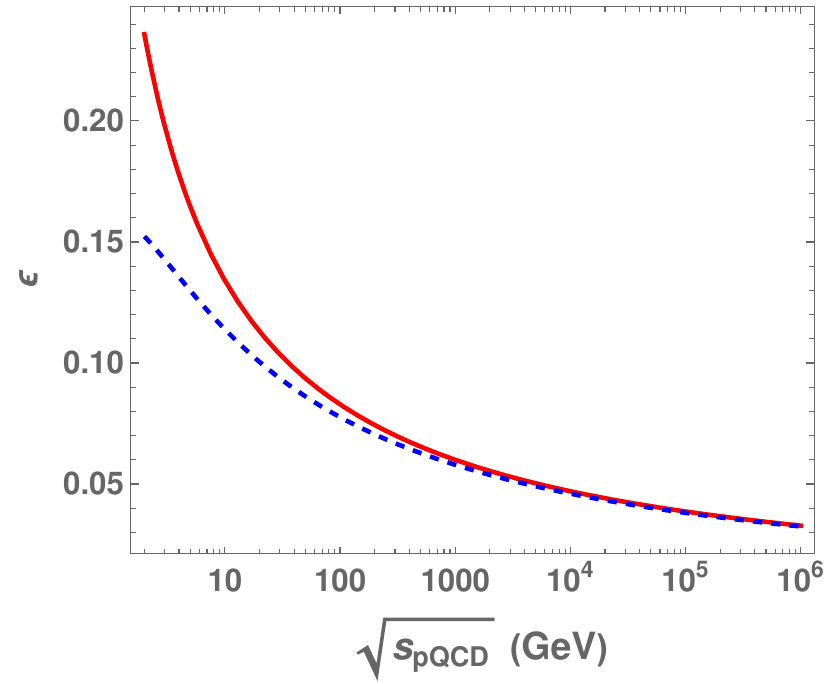}
\includegraphics[angle=0,width=.49\textwidth]{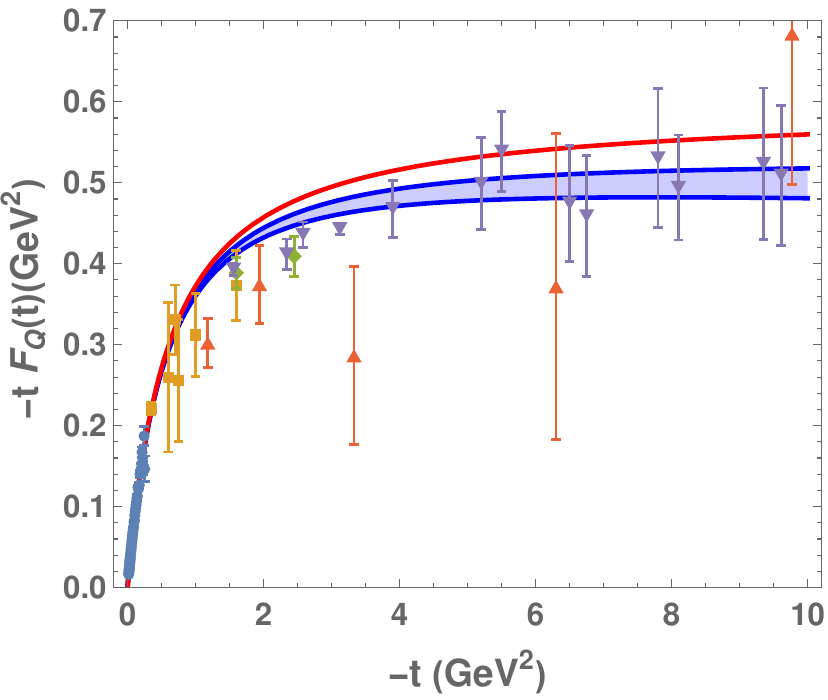}
\end{center}
 \caption{Left panel:  Regge-$\epsilon$ as a
 function of the matching scale with pQCD, $\sqrt{s_{\rm pQCD}}$; 
 Eq.~(\ref{eq:dlog}) (dashed) and asymptotic (solid).
  Right panel: Pion charge form factor from simple VMD 
 with $m_\rho=0.77$GeV compared with the FF obtained from Regge
 matching at the Regge scale $s_{R}= m_{\rho'}^2 $ and for
 $\epsilon=0.05-0.1$ (solid band). For experimental and lattice QCD data see Ref.~\cite{RuizArriola:2024gwb,Sanchez-Puertas:2024siv} and references therein.
\label{fig:ff}}
\end{figure}

An important application of the DR is to provide the FFs in
the space-like region, where the time-like region details become rather
irrelevant. In this state of afairs, the old meson dominance approach
becomes rather simple and reasonably accurate. From the QCD viewpoint,
it is based on the narrow character of hadronic resonances, an
assumption which is motivated by the large-$N_c$ limit. There, the width to mass ratio is $\Gamma/M
= {\cal O} (N_c^{-1}) \sim 1/3$, which is strongly supported by the
experimental Suranyi's average ratio $ \langle \Gamma/M \rangle_{\rm
PDG}= 0.12(18)$~\cite{Masjuan:2012gc}. The spectral function becomes a sum of {\it infinitely
many} narrow resonances, 
\begin{eqnarray}
\frac1\pi {\rm Im} F(s) = \sum_{i=0}^{\infty} c_{i} \delta (s-m_i^2) \implies
F(t) = \sum_{i=0}^\infty c_i \frac{m_i^2}{m_i^2 -t}. \label{eq:min}
\end{eqnarray}
The minimal hadronic antatz frequently used in the resonance physics is based on
saturating (\ref{eq:min}) with a {\it finite} number of narrow resonances (typically
one or two), disregarding the slow logarithmic scale running of
$\alpha_s$. The half-width rule can be used as a crude
estimate of
uncertainties~\cite{Masjuan:2012sk,Broniowski:2024oyk,Broniowski:2025ctl}. This
truncation only violates the highest SCSR.  A clear example is
provided by the charge FF of the pion, as given by the meson
dominance model, $F(-Q^2)= m_\rho^2 /(m_\rho^2+Q^2)$.  Schematically, SCSR becomes
\begin{equation} 
0 = \underbrace{\frac{1}{\pi} \int_{s_{\rm
th}}^{s_{\rm max}} ds \operatorname{Im} F (s)}_{m_\rho^2}
+ \underbrace{\frac{1}{\pi} \int_{s_{\rm max}}^{s_{\rm pQCD}}
ds \operatorname{Im} F (s)}_{-m_\rho^2 + \eta} +
\underbrace{  \frac{1}{\pi} \int_{s_{\rm pQCD}}^{\infty} \!\!\!\! ds \operatorname{Im} F (s) }_{-\eta} \, ,
\end{equation}  
where the first and the last term on the rhs represent the meson
dominance and the genuinely small pQCD contributions respectively.
Additional missing states (second term) are necessary to compensate. To fill the gap a 
model of the spectral density for $s_{\rm max} \le s \le s_{\rm pQCD}$, based on the radial Regge trajectories, $m_n^2 = a n +
b$~\cite{Masjuan:2012gc}, was proposed in
Ref.~\cite{RuizArriola:2025omi}.   In particular,
\[
\frac1\pi {\rm Im} F(s) = \begin{cases} \rho_{\rm ChPT} (s) &     4 m_\pi^2 \le s \le 16 m_\pi^2 \, , \qquad \text{ threshold region}\\
  \rho_{\rm R} (s) & 16 m_\pi^2 \le s \le \Lambda_R^2 \,
  , \qquad \text{isolated  resonance region}\\
\rho_{\rm Reg} (s) &     \Lambda_R^2 \le s \le \Lambda_{\rm pQCD}^2 \,  , \qquad \qquad \text{Regge region} \\ 
  \rho_{\rm pQCD} (s) & \Lambda_{\rm pQCD}^2 \le s \le \infty \,
  , \qquad \qquad \text{pQCD region} \end{cases}
\]
In the Regge region, $ s_{\rm max}=\Lambda_R^2 \leq
s \leq \Lambda_{\rm pQCD}^2$, the finiteness of the overlapping Radial
resonances with a fixed Suranyi's ratio $\Gamma_n/M_n \sim 0.12$ is
compatible with a behavior of the form $1/s^{1+\epsilon}$ for mesons
and $1/s^{2+2\epsilon}$ for baryons, which precludes the pQCD
$\alpha_s/s^{1,2}$ behavior.  The value of $\epsilon $ depends on the
matching point with pQCD, where the absolute normalization
is scale dependent and not very stable perturbatively. For that reason
we use a logarithmic derivative matching,
\begin{eqnarray}
\frac{d}{ds} \log \rho_{\rm Reg} (s)  |_{s=s_{\rm pQCD}} =  \frac{d}{ds} \log \rho_{\rm pQCD}(s) |_{s=s_{\rm pQCD}}
\label{eq:dlog}
\end{eqnarray}
which gives $ \epsilon \sim
1/\log(s_{\rm pQCD}/\Lambda_{\rm QCD}^2)$ asymptotically and yields
$\epsilon \sim 0.05-0.1$ for a very wide range of scales, see Fig.~\ref{fig:ff} (left), which can be
taken as a systematic error.  After all conditions are imposed, the
explicit pQCD and threshold pieces become tiny (see
e.g. \cite{RuizArriola:2025wyq,RuizArriola:2025omi}), and the FF reads
\begin{eqnarray}
F(-Q^2) = \sum_{n=1}^N c_n \frac{m_n^2}{m_n^2 + Q^2} + \frac{{\rm Im} F (s_R)}\pi \int_{s_R}^\infty ds \left(\frac{s_R}{s} \right)^{N + N\epsilon} \frac{1}{s+Q^2},
\end{eqnarray}
with $N_{\rm mes}=1$ and $N_{\rm bar}=2$.  The sum rules are used to fix $c_n$ and
${\rm Im} F (s_R)$. 

We illustrate the procedure for the charge pion form factor (see
also~\cite{RuizArriola:2024udm} for the gravitational FF).  In
addition to vector meson dominance with the $\rho$-meson we take the
Regge contribution from $\sqrt{s_R} = m_{\rho'}$, so that
\begin{eqnarray}
F_Q( t ) = \frac{ Z_\rho m_\rho^2}{m_\rho^2 -t } + \frac{{\rm Im} F_Q (s_R)}\pi \int_{s_R}^\infty ds \left(\frac{s_R}{s} \right)^{1+ \epsilon} \frac{1}{s-t},
\end{eqnarray}
The normalization and superconvergence conditions follow from
\begin{eqnarray}
1 = Z_\rho + \frac{{\rm Im} F_Q(s_R)}{\pi(1+\epsilon)} \, , \qquad 
0= Z_\rho m_\rho^2 + \frac{{\rm Im} F_Q(s_R) s_R}{\pi \epsilon}
\end{eqnarray}
respectively. Eliminating ${\rm Im} F_Q(s_R)$ and $Z_\rho$ we see that
the Regge contribution vanishes for $\epsilon \to 0$, which
corresponds to take $s_{\rm pQCD} \to \infty$. In
Fig.~\ref{fig:ff} (right) we can see the departure from simple VMD with
the mass value $m_\rho=0.77$GeV compared with the full FF for
$\epsilon=0.05-0.1$ and data.

\section{Conclusions}

Dispersive methods have a reputation of being rigorous and useful,
  until it comes to practical applications, where a high energy
  cut-off becomes explicit and influential. For hadronic form factors,
  most analyses ignore and/or violate the QCD superconvergence sum
  rules. The infinite radial Regge towers of states provide a simple
  framework where the smallness of the high energy tail becomes
  natural but not negligible. While the precise onset of
  pQCD is unknown our scheme provides a way to estimate this
  systematic uncertainty, which is rather competitive in the studied
  applications.


\begin{thebibliography}{10}

\bibitem{Masjuan:2012sk}
P. Masjuan, E. Ruiz~Arriola and W. Broniowski,
\newblock Phys. Rev. D 87 (2013) 014005, 1210.0760.

\bibitem{Lepage:1979zb}
G.P. Lepage and S.J. Brodsky,
\newblock Phys. Lett. B 87 (1979) 359.

\bibitem{Lepage:1979za}
G.P. Lepage and S.J. Brodsky,
\newblock Phys. Rev. Lett. 43 (1979) 545,
\newblock [Erratum: Phys.Rev.Lett. 43, 1625--1626 (1979)].

\bibitem{RuizArriola:2008sq}
E. Ruiz~Arriola and W. Broniowski,
\newblock Phys. Rev. D 78 (2008) 034031.

\bibitem{Ananthanarayan:2012tn}
B. Ananthanarayan, I. Caprini and I.S. Imsong,
\newblock Phys. Rev. D 85 (2012) 096006.


\bibitem{RuizArriola:2024gwb}
E. Ruiz~Arriola and P. Sanchez-Puertas,
\newblock Phys. Rev. D 110 (2024) 054003.

\bibitem{Sanchez-Puertas:2024siv}
P. Sanchez-Puertas and E. Ruiz~Arriola,
\newblock PoS QNP2024 (2025) 078.

\bibitem{RuizArriola:2025wyq}
E. Ruiz~Arriola, P. Sanchez-Puertas and C. Weiss,
\newblock Phys. Lett. B 866 (2025) 139585, 2503.10465.

\bibitem{RuizArriola:2025omi}
E. Ruiz~Arriola and P. Sanchez-Puertas,
\newblock Eur. Phys. J. ST  (2026), 2507.22726.

\bibitem{Cao:2025dkv}
X.H. Cao et~al.,
\newblock Eur. Phys. J. ST  (2025), 2507.05375.

\bibitem{Hoferichter:2016duk}
M. Hoferichter et~al.,
\newblock Eur. Phys. J. A 52 (2016) 331, 1609.06722.

\bibitem{Masjuan:2012gc}
P. Masjuan, E. Ruiz~Arriola and W. Broniowski,
\newblock Phys. Rev. D 85 (2012) 094006, 1203.4782.

\bibitem{Broniowski:2024oyk}
W. Broniowski and E. Ruiz~Arriola,
\newblock Phys. Lett. B 859 (2024) 139138.

\bibitem{Broniowski:2025ctl}
W. Broniowski and E. Ruiz~Arriola,
\newblock Phys. Rev. D 112 (2025) 054028.

\bibitem{RuizArriola:2024udm}
E. Ruiz~Arriola and W. Broniowski,
\newblock PoS QNP2024 (2025) 068.

\end{thebibliography}

\end{document}